\documentclass{imsart}
\RequirePackage{natbib}

\usepackage{latexsym,amssymb, amsthm, epsfig}

\usepackage{color}
\usepackage{graphicx}
\usepackage{amsmath}
\usepackage{amsfonts}
\usepackage{bbm}
\usepackage{psfrag}

\newtheorem{theorem}{\bf Theorem}[section]
\newtheorem{lemma}{\bf Lemma}[section]
\newtheorem{proposition}{\bf Proposition}[section]
\newtheorem{remark}{\bf Remark}[section]
\newtheorem{definition}{\bf Definition}[section]

\def\S{{\cal S}}

\def\R{{\mathbb R}}

\def\A{{\mathcal{A}}}

\def\C{{\mathcal{C}}}

\def\qed{\hfill $\blacksquare$}
\def\({{\Bigl(}}
\def\){{\Bigr)}}

\def\ind{{\mathbf 1}}
\def\square{\ifmmode\sqr\else{$\sqr$}\fi}
\def\sqr{\vcenter{
         \hrule height.1mm
         \hbox{\vrule width.1mm height2.2mm\kern2.18mm\vrule width.1mm}
         \hrule height.1mm}}                  % This is a slimmer sqr.

\theoremstyle{plain}

\theoremstyle{definition}

\theoremstyle{remark}

\def\ep{\end{proof}}

\def\bp{\begin{proof}}

\def\epsilon{\varepsilon}

\psfrag{infini}{$\infty$}
\psfrag{c1}{$c_1$}
\psfrag{c2}{$c_2$}
\psfrag{r1}{$r_1$}
\psfrag{r2}{$r_2$}
\psfrag{r3}{$r_3$}
\psfrag{rho1}{$\rho_1$}
\psfrag{rho2}{$\rho_2$}
\psfrag{reg1}{$r_1 = 0$}
\psfrag{reg2}{$r_1 > 0$}
\psfrag{b1}{$C$}
\psfrag{infini}{$\infty$}
\psfrag{rho1 = 0.6, K = 1000}{\hspace{-2mm}$\scriptscriptstyle\rho_1 = 0.6, K = 1000$}
\psfrag{rho1 = 0.6, K = 10000}{\hspace{-2mm}$\scriptscriptstyle\rho_1 = 0.6, K = 10000$}
\psfrag{rho1 = 0.6, K = infini}{$\scriptscriptstyle\rho_1 = 0.6, K = \infty$}
\psfrag{rho1 = 0.5, K = 1000}{\hspace{-2mm}$\scriptscriptstyle\rho_1 = 0.5, K = 1000$}
\psfrag{rho1 = 0.5, K = 10000}{\hspace{-2mm}$\scriptscriptstyle\rho_1 = 0.5, K = 10000$}
\psfrag{rho1 = 0.5, K = infini}{$\scriptscriptstyle\rho_1 = 0.5, K = \infty$}
\psfrag{rho1 = 0.7, K = 1000}{\hspace{-2mm}$\scriptscriptstyle\rho_1 = 0.7, K = 1000$}
\psfrag{rho1 = 0.7, K = 10000}{\hspace{-2mm}$\scriptscriptstyle\rho_1 = 0.7, K = 10000$}
\psfrag{rho1 = 0.7, K = infini}{\hspace{-2mm}$\scriptscriptstyle\rho_1 = 0.7, K = \infty$}
\psfrag{rho1 = 0.25, K = 1000}{\hspace{-3mm}$\scriptscriptstyle\rho_1 = 0.25, K = 1000$}
\psfrag{rho1 = 0.25, K = 10000}{\hspace{-4mm}$\scriptscriptstyle\rho_1 = 0.25, K = 10000$}
\psfrag{rho1 = 0.25, K = 1000000}{\hspace{-5mm}$\scriptscriptstyle\rho_1 = 0.25, K = 1000000$}
\psfrag{rho1 = 0.25, K = infini}{\hspace{-2mm}$\scriptscriptstyle\rho_1 = 0.25, K = \infty$}
\psfrag{rho1 = 0.3, K = 1000}{\hspace{-3mm}$\scriptscriptstyle\rho_1 = 0.3, K = 1000$}
\psfrag{rho1 = 0.3, K = 10000}{\hspace{-4mm}$\scriptscriptstyle\rho_1 = 0.3, K = 10000$}
\psfrag{rho1 = 0.3, K = 1000000}{\hspace{-5mm}$\scriptscriptstyle\rho_1 = 0.3, K = 1000000$}
\psfrag{rho1 = 0.3, K = infini}{$\scriptscriptstyle\rho_1 = 0.3, K = \infty$}
\psfrag{K = 1000}{\hspace{-3mm}$\scriptscriptstyle K = 1000$}
\psfrag{K = 10000}{\hspace{-2mm}$\scriptscriptstyle K = 10000$}
\psfrag{K = infini}{$\scriptscriptstyle K = \infty$}

\def\ep{\hfill $\Box$}

\def\proof{\noindent{\bf Proof.}\ }
\def\bp{\noindent{\it Proof.}\ }
\def\ep{\hfill $\Box$}

\newcommand{\Ints}{\ensuremath{\mathbb{N}}}
\newcommand{\bN}{\ensuremath{\bar{\Ints}}}
\renewcommand{\ind}{\ensuremath{\mathbbm{1}}}
\newcommand{\mean}{\ensuremath{\mathbb{E}}}

\newcommand{\bS}{\ensuremath{\bN^{N-1}}}
\newcommand{\diff}{\mathop{}\mathopen{}\mathrm{d}}
\renewcommand{\Pr}{\ensuremath{\mathbb{P}}}

\arxiv{math.PR/0000000}

\begin{document}

\begin{frontmatter}

\title{Bandwidth sharing networks\\ with priority scaling}
\runtitle{Bandwidth sharing networks with priority scaling}
\begin{aug}
  \author{\fnms{M.}  \snm{Feuillet}\ead[label=e1]{mathieu.feuillet@inria.fr}},
  \author{\fnms{M.} \snm{Jonckheere}\ead[label=e2]{mjonckhe@dm.uba.ar}}
  \and
  \author{\fnms{B. J.}  \snm{Prabhu}%
  \ead[label=e3]{bala@laas.fr}}

  \runauthor{M. Feuillet et al.}

  \affiliation{INRIA, CONICET and CNRS, LAAS}

  \address{INRIA Paris-Rocquencourt,\\
           Domaine de Voluceau, Rocquencourt, B.P. 105\\
           78153 le Chesnay Cedex, France.\\
           \printead{e1}}

  \address{CONICET, Departamento de Matem\'atica,\\
           Facultad de Ciencias Exactas y Naturales,\\
           Universidad de Buenos Aires,\\
           Pabell\'on 1, Ciudad Universitaria,\\
           1428 Buenos Aires, Argentina.\\
           \printead{e2}}

  \address{CNRS; LAAS;\\ 7 avenue du Colonel Roche,\\ F-31077 Toulouse, France.\\
          Universit\'e de Toulouse;\\ UPS, INSA, INP, ISAE; LAAS;\\ F-31077 Toulouse, France.\\
          \printead{e3}}
\end{aug}

\begin{abstract}
In multi-class communication networks, traffic surges due to one
class of users can significantly degrade the performance for other
classes. During these transient periods, it is thus of crucial
importance to implement priority mechanisms that conserve the
quality of service experienced by the affected
classes, while  ensuring that the temporarily unstable class is not
entirely neglected. In this paper, we examine the complex interaction
occurring between several classes of traffic when classes obtain bandwidth
proportionally to their incoming traffic.
 We characterize the evolution of the
network from the moment the initial surge takes place until the
system reaches its equilibrium. Using an appropriate scaling,
we show that the trajectories of the temporarily
unstable class can be described by a differential equation, while
those of the stable classes retain their stochastic nature.
A stochastic averaging phenomenon occurs and the dynamics of the temporarily
unstable and the stable classes continue to influence one another.
We further proceed to characterize the obtained differential
equations and the stability region under this scaling for monotone networks.
We illustrate these result on several toy examples
and we finally build a penalization rule using these results for
a network integrating streaming and elastic traffic.
\end{abstract}



\end{frontmatter}

\section{Introduction}

Communication networks are dealing with
heterogeneous sources of traffic having
different behaviors in terms of volume of data and aggressivity.
Ideally, the network should respond to the different demands in the fairest possible way, i.e. by avoiding a significant degradation of quality of service of a given class of traffic when
another class undergoes a major traffic surge.

The impact of large-scale traffic surges, also known as
flash-crowds, on web servers and content
distribution networks has been the subject of several studies
\citep{SRS04, KKJB05, DMCVM07}. These mainly focus on designing
mechanisms to make the content providers resilient to surges {of a
given type of traffic}. However, in addition to overloading the
content providers, a traffic surge can also negatively impact the
performance of concurrent flows in the network. The
temporarily unstable class can potentially starve the other classes
from network capacity thereby subjecting them to unreasonable delays
and packet losses. In such circumstances, in addition to protection
mechanisms in web servers, it is crucial to implement
bandwidth-sharing mechanisms inside the network that would protect
the stable classes from the adversarial effects of the surge. It
seems natural that such mechanisms should penalize the temporarily
unstable class more when the level of congestion it creates is
larger, without actually throttling it. (Thus, the more significant
the surge is, the smaller the bandwidth each flow in this class
gets.) The consequences of traffic surges on the performance of the
different classes in the presence of such bandwidth sharing
mechanisms have not been explored much.

In this paper, we take a global view at the effects of a traffic surge
in a multi-class communication network: our aim is to present an analytic
treatment of the complex interaction that takes place between the temporarily
unstable class and the stable class during a traffic surge when the
temporarily unstable class is penalized proportionally to its level of
congestion.

Towards this end, we consider stochastic networks describing the evolution
of the number of flows in a network where different
classes of traffic compete for the bandwidth. Bandwidth-sharing network
models~\citep{massrob,bp,gromoll} have become quite a standard modeling
tool over the past decade. In particular,
they have been used extensively to represent the flow level dynamics of data
traffic in wireline or wireless networks \citep{BFreview}, as well as for the
integration of voice and data traffic \citep{bonaldsig2004}, hence generalizing
more traditional voice traffic models, e.g.~\citep{kelly1979}.

To obtain structural results, we introduce a scaling when possibly only a subset of classes
have initial conditions converging to infinity. Those classes shall be the ones undergoing a traffic surge (surging classes). We consider a situation where the allocation of bandwidth shall be meanwhile weighted such that
the other classes of traffic (stable classes) are not led to starvation, i.e., the priority weight is very small compared to the offered traffic.
Accelerating time together with
re-scaling the state of the surging classes allow to ``zoom out''
the process, just as for usual fluid limits and obtain a bird's-eye
view of the large scale dynamics for these classes.
 In order to obtain a classical fluid limit for Jackson networks
\citep{robert} or for more complex bandwidth-sharing networks \citep{gromoll}, all
the classes are {jointly} scaled in time and in space. This yields a set of differential
equations that govern the dynamics of all the classes. Under additional assumptions
on the drift $\delta$ of the considered Markov process, the differential equation is
simply of the form ${\dot x}(t)=\delta(x(t))$ (see the considerable amount
of work on fluid limits and ODE methods both for Markov processes and for communications networks \citep{dai95,darling,gromoll,meynbook,robert}).

In our case, the situation differs as the transitions of surging classes are also scaled
to model that the priority weight of surging classes is inversely proportional to the level
congestion.
This has far-reaching consequences for the structure of the limiting process.
Under this scaling, we will show that the dynamics of the surging classes can be
described by a set of deterministic differential equations, while the
stable classes retain their stochastic nature. Hence, a time-scale separation
occurs: the surging classes evolve on the much slower time-scale
compared to the stable classes.
However even with this separation of time-scales, a strong coupling in the dynamics
of the surging classes and the stable classes remains. The dynamics of the
surging classes are influenced by the stable classes through their conditional
distribution which in turn depends on the level of congestion of surging-class flows being
fixed to their present macroscopic value.  Hence, for the surging classes the differential equations obtained are of the form
${\dot x}(t)=\bar \delta^t(x(t))$, where $\bar \delta^{t}$ is an average of
the first coordinates drift according to the conditional distribution of the other classes,
given the state of the surging classes. This phenomenon is usually known in the probability literature as  averaging principle and has been studied by several authors.
We follow in particular the methodology introduced in the seminal paper \citep{kurtz}.
In the analysis of the fluid limit of bandwidth sharing networks
a time-scale separation between classes usually occurs when one class of traffic
reaches equilibrium faster than the others, and hence when the fluid limit hits an hyperplane of the state space.
Simple examples of this phenomenon can be found in \citep{robert}. A more complex
example can be found in \citep{mathieu2010}.
The interesting feature in our scaling is the appearance of the averaging principle
in the whole state space.
 Similar averaging phenomena have also been studied in statistical physics \citep{kipnis1991}
as well as chemistry and biochemistry \citep{segel} where
the kinetics of chemical reactions can be described by systems of ordinary differential equations.
 Usually these works assume that one of the dependent variable
is in steady state with respect to the instantaneous values of the other dependent variables. Taking this time-scale separation as an assumption, an efficient approximation method called the quasi-steady-state is commonly used in that context.
This is however in contrast with our situation where we show that the time-scale decoupling
occurs as a consequence of the scaling of the parameters of the transitions of the stochastic processes considered.

\subsection*{Contribution:}

Our contribution first consists in establishing the convergence in $L^1$ uniformly on compact sets  for stochastic processes commonly adopted in the modeling and analysis of communications networks under the scaling considered. Since the slow part of the processes (surging classes) remains coupled (at a macroscopic scale) to the fast part (the remaining classes), such a proof is not standard and has to be decomposed in several steps. While preliminary results for monotone networks were presented in \citep{itc22}, a general proof for general bandwidth sharing networks was still missing.

Second, we characterize the responses (evolutions of queue length) of different networks to the
surge of traffic.
We introduce the notion of robust stability, which describes a situation when the network
can resorb a surge of traffic by eventually reducing the macroscopic state of
all classes to $0$.
We call the set of traffic parameters that lead to this condition the
robust stability region.
We characterize this robust stability region for work conserving allocations and for monotone allocations.
 We first show that for work conserving allocations, the unstable class, at its macroscopic time scale, ``sees'' the other classes as having full priority, while the effect of the first class on the other classes gradually vanishes (again at a macroscopic time scale). Hence surging classes tend macroscopically to $0$ under the stability condition of the system  $\sum_{j=1}^N \rho_j< 1 $. The situation is
more complex for non work-conserving networks, where the behavior of the unstable class
depends in an intricate manner upon that of the other classes. In particular, under the usual
stability conditions of the network, the macroscopic state  of \mbox{surging classes} might converge to $0$ or to a strictly positive number, depending on the conditional distribution of the other classes. For monotone networks, we prove that robust stability
boils down to stability under an allocation giving full priority to stable classes.
We illustrate these concepts on several simple network topologies.
Finally, we use our analytical results to build an implementable penalization rule allowing to adapt the level of priority of streaming traffic in a network integrating streaming and elastic traffic, such as to target a given loss probability threshold/ quality of service.

The rest of the paper is organized as follows. The model is presented in next section.
In Section~\ref{sec:fluid}, we present the convergence theorem for the considered scaling.
In Section~\ref{sec:qual}, we analyze the qualitative behavior of networks after a traffic
surge in different cases and we give numerical examples of applications of the main result
to bandwidth sharing on some simple network topologies.
In Section~\ref{sec:integration}, we construct a practical penalization rule for streaming traffic. Finally, we conclude in Section~\ref{sec:conclusion}.

\section{Model }\label{sec:main}

\subsection*{Notation}

In the sequel, for $x \in \mathbb Z^N$, $|\cdot |$ denotes the $l_1$-norm:
$$|x|=\sum_{i=1}^N |x_i|.$$
For $x,y \in \mathbb Z^N$, we also use the notation $x \le y$ to denote the partial order $x_i \le y_i$ for all $i=1 \ldots N$.

\subsection{Networks with traffic weighted allocations}
We consider a processor sharing network with $N$ traffic classes.
Within each of the $N$ traffic classes, resources are shared according to a processor-sharing service
discipline. The service rates are state-dependent: they may depend
on the number of flows within the same class, as well as on the
numbers of flows in all other classes. The service rates of the $N$
traffic classes will be denoted by $\phi=(\phi_i(\cdot))_{i=1}^N$.
Several examples are considered in the next
section. Note that the service rate function $\phi$ captures the
allocation of bandwidth which is determined by the specific network
topology and congestion control mechanisms. Special
allocation functions that have received much attention in literature
include the celebrated max-min fair allocation and the proportional
fair allocation.

We assume that class-$i$ customers arrive subject to a Poisson
process of intensity $\lambda_i$ and require exponentially
distributed\footnote{Such assumptions are certainly not necessary to
obtain the results we are aiming at; however, a rigorous
generalization would be technically very involved and is beyond the
scope of the present paper.} service times of mean $\mu_i^{-1}$ for
class-$i$. The arrival processes of all classes are mutually
independent. Our main results allow for time-varying arrival rates
for the class exhibiting a traffic surge. When applicable, we
reflect this dependence in the notation by adding the time parameter to the
arrival rates and then $\lambda_1(t)$ is the arrival rate of class-$1$
at time $t$. For ease of exposition, however, we restrict
ourselves to constant arrival rates for all classes in this section
and will formulate our results with time-varying arrival rates in
Section~\ref{sec:main}.

Let $X$ be the stochastic process
describing the number of flows (or calls) in progress.
In the absence of priority mechanisms, and under the assumptions of Poisson
arrivals and exponential flow sizes, $X$ is a multi-dimensional birth
and death process with transition rates:
\begin{align*}
q(x,x-e_i)&=\mu_i \phi_i(x),\\
q(x,x+e_i)&=\lambda_i,
\end{align*}
with $x \in \mathbb N^N$.
Assume now that priority mechanisms are employed in the network
such that the actual bandwidth allocation depends on the variables
$r_i x_i,~ i=1\ldots N$ rather than simply on $x_i,~ i=1\ldots N$.
Hence, if $x_i$ is thought  of as a measure of the level of congestion
of class-$i$, a differentiation between classes can be enforced by
giving different weights to the different classes. (Such a differentiation
can be enforced at lower time-scales by packet schedulers like
weighted deficit round robin.)

It can also be the case that each class of traffic has a limited
peak rate (because of access constraints for instance). It could then be
advantageous for providers, in order to meet the demand, to share
capacity as a function of  the demanded rates $r_i x_i$ rather than
as a function of the number of flows of each class in the network.
In both configurations, $X$ can now be described as multi-dimensional birth
and death process with transition rates:
\begin{align*}
q(x,x-e_i)&=\mu_i \phi_i(r.x),\\
q(x,x+e_i)&=\lambda_i,
\end{align*}
where $r.x=(r_i x_i)_{i=1 \ldots N}$ for some $r \in \mathbb R^N_+$.
To avoid confusion, we emphasize once more that reflecting the
dependence on the control parameters $r_i$ in our notation will be
more convenient for the purposes in this paper, rather than making
this dependence implicit, i.e., through the allocation $\tilde
\phi(x)=\phi(r.x)$.
The load of class $i$ is given by
$$\rho_i={\lambda_i \over \mu_i}.$$
We shall further suppose that the weights $r$ are chosen for each class proportionally
to the size of the traffic surge, i.e.,
$$r_i=r_i (|x|)={ \omega_i \over |x|}.$$
Finally, denote the global load of the system by $\bar \rho=\sum_{i=1}^N \rho_i$.

\section{Fluid limits with time scales decoupling}\label{sec:fluid}

We model a traffic surge by a large number of initial flows and large arrival rates
for a subset of classes being (temporarily at least) unstable.
Let $c \le N \in \mathbb N$.
To get structural results on the process $X$, we study the case where:
\begin{enumerate}
\item
the number of initial class-$i$, $i=1 \ldots c$ flows is of order $K=|x|$.

\item we scale (accelerate) time by a factor $K$,

\item we scale class-$i$, $i=1\ldots c$ states by a factor $1/K$,

\item
the prioritization weight $r_i$ of class-$i$, $i \le c$ is of order $1/K$.
\end{enumerate}

We now consider a network with several classes of traffic and with
class-$i$, $i=1 \ldots c$  going through a temporary surge of traffic. Recall that we
focus on a regime where $r_i\equiv {\omega_i \over K}$ and $K \to \infty$.
We further let $Y^K$ denote the (scaled) process:
\begin{equation}\label{eq:Y}
Y^K(t)= \left(({X^K_i(Kt) \over K})_{i=1 \ldots c}, (X^K_i(Kt))_{i=c+1 \ldots N} \right).
\end{equation}
In the following we show that, as $K\to\infty$, $Y^K$ converges to a
stochastic process with a deterministic first coordinate, which is a
solution of a differential equation which we describe in terms of an
averaged rate $\bar \phi$. In the limit, the result implies a time-scale
separation between the first classes and the other ones.

Define $U^{z} $ to be a $N-c$ dimensional Markov birth-and-death
process with arrival rates $\lambda_i$ and death rates
$\phi_i(z,\cdot)$, $i=c+1\ldots N$ ($z \in \mathbb R^k_+$) and denote by $\pi^{z}(\cdot)$ its stationary probability (when it exists).
 When we do not use a time index, we implicitly suppose
that we consider stationary versions of the processes.

For a given $a(t)\in\R^k$, let $u(t) \in \mathbb R^k$ be the solution (assuming it  exists and it is unique) of the differential equation:
\begin{equation}\label{eq:diff}
\forall i=1\ldots c,  \ \ \dot{u_i}(t)=
\begin{cases}
{\dot a_i}(t)- \bar \phi_i(u(t)), &\text{if } u_i(t)>0,\\
0, &\text{if } u_1(t)=0,
\end{cases}
\end{equation}
with $\bar \phi_i(z)=\sum_{y \in \mathbb N^{N-c}}    \phi_i(z, y) \pi^{z}(y).$
To establish our main result, we shall make the following assumptions:
\begin{description}
\item[$(A_1)$:] $\phi_i(\cdot, x_{c+1},\ldots,x_N)$ can be extended to Lipschitz-continuous functions from $\mathbb R^{c}_+\setminus\{0 \}$ to $\mathbb R^+$.
\item[$(A_2)$:] for all fixed $z$, the process $U^{z}$ is ergodic.
We can thus define $\mean^{U^{z}}$ the mean under the stationary distribution of
the process $U^{z}$ .
\item[$(A_3)$:] ${1 \over K} \int_{0}^{Kt} \lambda^K_i(s)\diff s \to  a_i(t), \ i=1\ldots c.$
\end{description}

We can now proceed to state our main result:

\begin{theorem}\label{theo:thetheo}
 Under the assumptions $(A_1)$, $(A_2)$ and $(A_3)$, the process
  ${Y_i^K(t)}_{i=1\ldots c}$ converges in $L^1$, uniformly on compact intervals, to the deterministic trajectory $u(t)$, i.e.,
\begin{equation}\label{eq:theo1}
\mean\left[\sup_{0 \le s\le t}\left|{Y_i^K(s)} - u_i(s)\right|\right]  \to 0, ~ ~  K \to \infty, \ \forall i=1\ldots c.
\end{equation}
Moreover, for all times $t$, and for all bounded continuous functions $f$:
\begin{equation}
\begin{aligned}
\lim_{K \to \infty}\mean \Biggl( \sup_{0\leq s \leq t} &\Biggl| \int_0^s f\left(Y^{K}(u)\right)\\
&- \mean^{U^{Z(s)}}\left( f\left(Z(s),U^{Z(s)}(s)\right) \;\middle|\; Z(s)=u(s) \right) \diff s \Biggr| \Biggr) = 0.
\end{aligned}
\label{eq:averaging}
\end{equation}
\end{theorem}

The details of the proof are given in the next Section.
We underline here the main steps:
\begin{itemize}
\item
We first prove tightness of the laws of the scaled process, and show that the limit-points of $Y_1^K$ are continuous processes.
\item
Supposing the convergence in distribution of the first class
we characterize the limit of the functional $ \int_0^t \ind_{\{(X^K_2(Ks),\dots,X^K_N(Ks)) \in \Gamma \}}\diff s,\ \forall \Gamma \subset \bN^{N-1}$ and prove the limits are unique (and deterministic given the value of the first class).
A key step is the useful characterization of bimeasures.
\item
Finally, we show that $Y^K_1$ converges in distribution towards a deterministic process which allows to prove, using the previous step, the convergence in $L^1$, uniformly on compact sets.
\end{itemize}

\subsection{Proof of Theorem \ref{theo:thetheo}}
\label{sec:proof-averaging}

\subsubsection*{Step 1:}

For the ease of exposition,  we suppose that class-$1$ only undergoes
a surge of traffic. The proof then extends directly to the general case.

We thus consider the process $Y^K(t)= \left({X^K_1(Kt) \over K}, (X^K_i(Kt))_{i=2 \ldots N} \right)$
as defined by \eqref{eq:Y}. We define $\bN = \Ints \cup \{+\infty\}$ and for each $K$,
we define the following random measure on $[0,\infty)\times\bN^{N-1}$:
$$
\nu^K((0,t)\times \Gamma) =  \int_0^t \ind_{\{(X^K_2(Ks),\dots,X^K_N(Ks)) \in \Gamma \}} \diff s,\
\forall \Gamma \subset \bN^{N-1},\text{ and } \forall t\geq0.
$$
We denote $\mathcal{L}_0(\bN^{N-1})$ the set of measures on $[0,\infty)\times\bN^{N-1}$
such that, for all measure $\nu$ in $\mathcal{L}_0(\bN^{N-1})$ and all $t\geq0$,
we have $\nu((0,t)\times\bN^{N-1}) = t$. Since $\bN$ is compact, we have that
$\mathcal{L}_0(\bN^{N-1})$ is compact and we deduce that
$\{\nu^K\ K\in\Ints \}$ is relatively compact.

In order to prove the relative compactness of $\{(Y^K_1,\nu^K),\ K\in\Ints \}$,
we then just have to prove the relative compactness of $\{Y^K_1,\ K\in\Ints \}$.
We define the following process
\begin{equation}
M_1^K(t) = Y_1(t) - \frac{1}{K} \int_0^{Kt} \lambda_1(s) \diff s
+ \frac{1}{K} \int_0^{Kt} \phi_1\left(\frac{X^K_1(s)}{K},X^K_i(s)\right)\diff s
\label{eq:M}
\end{equation}
The martingale characterization of jump processes (see \citep{rogers-00}) shows
that $M_1^K$ is a locale martingale and its increasing process is given by
$$
\langle M_1^K \rangle  = \frac{1}{K^2} \int_0^{Kt} \lambda_1(s)\diff s
+  \frac{1}{K^2} \int_0^{Kt} \phi_1\left(\frac{X^K_1(s)}{K},X^K_i(s)\right)\diff s
$$
Using Doob's inequality \footnote{For any martingale $M$, using Cauchy Schwartz and Doob's inequality \citep{darling}, we get that:
\begin{align*}
\mean \left( \left| \sup_{0\le s \le t} M_s\right|  \right)^2 &\leq \mean\left(  \sup_{0\le s \le t} |M_s|  \right) ^2 \\
 &\leq  \mean \left(  \sup_{0\le s \le t} M_s^2  \right),\\
   &\leq 4 \mean \left(  M_t^2  \right).
 \end{align*}}, it follows that $M_1^K$ converges in probability to 0 on any compact set
when $K\to\infty$, i.e., for any $T\geq 0$ and any $\varepsilon>0$,
\begin{equation}
\lim_{K\to\infty} \Pr\left( \sup_{0\leq s \leq t} |M^K(s)| > \varepsilon \right) =0.
\label{eq:M-conv}
\end{equation}
We then define $w_h$ the modulus of continuity for any function $h$
defined on $[0,t]$:
$$
w_h(\delta) = \sup_{s,u\leq t;\ |u-s|<\delta} |h(s)-h(u)|.
$$
Using Equations \eqref{eq:M} and \eqref{eq:M-conv}, we are able to prove that
for any $\varepsilon>0$ and $\eta >0$, there exists $\delta>0$ and $A$ such that
for $K>A$, we have
$$
\Pr \left ( w_{Y^K_1(.)}(\delta) > \eta \right) \leq \varepsilon.
$$
The conditions of \cite[7.2 p81]{billingsley-99} are then fullfilled and the
set $\{Y^K_1,\ K\in\Ints \}$ is relatively compact. Moreover, any limiting
point is a continuous process.

\subsubsection*{Step 2:}

We now suppose that $(Y^K_1)$ converges in distribution to a limit $Z_1$.
We have to characterize any limiting point of the sequence $(\nu^{K})$
and then deduce the existence and uniqueness of the limit of $(\nu^{K})$.
In the following, we consider a convergent subsequence $(Y^{K_l},\nu^{K_l})$ and
its limit process $(Z_1,\nu)$.

Let $(\Omega,\mathcal{F},\Pr)$ be the probability space on which they are defined.
We call $\{\mathcal{F}_t\}$ the natural filtration of $(Z(t),\nu)$.
We then define $\gamma$ such that
$$
\forall A \in \mathcal{F},\ \forall B \in
\mathcal{B}([0,\infty)), \forall C \in \mathcal{B}(\bS)\ \gamma(A\times B\times C) =
\mean(\ind_A\nu(B \times C)).
$$

According to \cite[appendix 8]{ethier-86}, $\gamma$ can be extended to a
measure on
$\mathcal{F}\otimes\mathcal{B}([0,\infty))\otimes\mathcal{B}(\bS)$ and there exists
$\vartheta$ such that for all $t$, $\vartheta(t,.)$ is a
random probability measure on $\bS$
and for any $B\in \mathcal{B}(\bS)$, $(\vartheta(t,B), t\geq 0)$ is
$\{\mathcal{F}_t\}$-adapted and for any $A\in\mathcal{F}\otimes\mathcal{B}([0,\infty))$,
\begin{equation}
  \gamma(A\times B)
  =\mean \left( \int_0^{+\infty} \ind_A(s)\vartheta(s,B)\diff s
  \right).
\label{eq:modiano}
\end{equation}
$$
M_B(t) = \nu([0,t]\times B) - \int_0^t \vartheta(s,B)\diff s.
$$
$M_B$ is $\{\mathcal{F}_t\}$-adapted and continuous.
We consider $t\geq s$, and $D\in \mathcal{F}_s$. We define
$\ind_C(\omega,u)=\ind_D(\omega)\ind_{[s,t)}(u)$ and we have
\begin{align*}
  \mean\left( \ind_D \nu([s,t)\times B) \right) &= \gamma(D\times[s,t)\times B),\\
                                                      &= \gamma(C\times B),\\
                                                      &= \mean\left(\int_0^\infty \ind_C(u) \vartheta(u,B)\diff u\right),\text{ (according to \eqref{eq:modiano})}\\
  &= \mean\left( \ind_D \int_s^{t} \vartheta(u,B)\diff u\right).
\end{align*}
Since the previous equality is true for all $D\in\mathcal{F}_t$, it follows that
$$
\mean\left( \nu([s,t)\times B) \;|\; \mathcal{F}_s \right) = \mean\left(
  \int_s^{t} \vartheta(u,B)\diff u \;\middle|\; \mathcal{F}_s \right).
$$
and immediatly, we have
$$
\mean\left(M_B(t) \;\middle|\; \mathcal{F}_s\right) = M_B(s).
$$

Then, $M_B$ is a continuous $\{\mathcal{F}_t\}$-martingale. It
has finite sample paths and then is almost surely identically
null. Almost surely, the following equation holds for all $t$,

\begin{equation}
\forall B\subset \bS,\ \nu([0,t)\times B) = \int_0^t \vartheta(s,B)\diff s.
\label{eq:vartheta}
\end{equation}

We have to characterize the random measures $\vartheta(t,.)$ associated to $\nu$.
For any uniformly continuous bounded function $g$ on $\bS$ and any $K\in\Ints$, we define
\begin{align*}
   M^K_g(t) = \frac{1}{K} \Bigl(g\bigl(X^K_2&(Kt),\dots,X^K_N(Kt)\bigr) - g(0)\Bigr)\\
  - \sum_{i=2}^N\lambda_i \int_0^t &\Bigl(g\left(X^K_2(Kt),\dots,X^K_i(Kt)+e_i,\dots,X^K_N(Kt)\right)\\
           &- g\left(X^K_2(Kt),\dots,X^K_N(Kt)\right) \Bigr) \diff s  \\
   - \sum_{i=2}^N \mu_i \int_0^t &\Bigl(g\left(X^K_2(Kt),\dots,X^K_i(Kt)-e_i,\dots,X^K_N(Kt)\right)\\
                     & - g\left(X^K_2(Kt),\dots,X^K_N(Kt)\right)\Bigr)\\
& \phi_i\left(Y_1^K(s),X^K_2(Kt),\dots,X^K_N(Kt)\right)\diff s.
\end{align*}

As $M^K$ is a martingale, $M^K_g$ is a martingale. We have that
$M^{K_l}_g$ converges in distribution to $0$. $|K_l|^{-1}
(g(X^{K_l}_2(K_lt),\dots,X^{K_l}_N(K_lt)) - g(0))$ also converges to $0$ because
$g$ is bounded. As a consequence, the following term
\begin{align*}
  \sum_{i=2}^N \lambda_i \int_0^t
  \biggl(&g\left(X^{K_l}_2(Kt),\dots,X^{K_l}_i(K_lt)+e_i,\dots,X^{K_l}_N(K_lt)\right)\\
   & - g\left(X^{K_l}_2(K_lt),\dots,X^{K_l}_N(K_lt)\right)\biggr)\diff s  \\
  -\sum_{i=2}^N \mu_i \int_0^t & \biggl( g\left(X^{K_l}_2(K_lt),\dots,X^{K_l}_i(K_lt)-e_i,\dots,X^{K_l}_N(K_lt)\right)\\
           & - g\left(X^{K_l}_2(K_lt),\dots,X^{K_l}_N(K_lt)\right)\biggr)\\
          &\phi_i\left(Y_1^{K_l}(s),X^{K_l}_2(K_lt),\dots,X^{K_l}_N(K_lt)\right)\diff s
\end{align*}
also converges in distribution to $0$. But, by the continuous mapping theorem and
\eqref{eq:vartheta}, it converges in distribution to
\begin{align*}
\int_0^t \sum_{i=2}^N \Biggl( &\lambda_i \sum_{y \in \Ints^{N-1}} g(y+e_i) -
  g(y)\\ &+ \mu_i \sum_{y \in \Ints^{N-1}} \left(g(y-e_i) - g(y)\right)
  \phi_i\left(Z_1(s),y\right)\Biggr) \vartheta(s,y)\diff s.
\end{align*}
Consequently, this is null almost surely for all $t$ and we have then, for
Lebesgue-almost every $t$,
\begin{align*}
\sum_{i=2}^N \Biggl( &\lambda_i \sum_{y \in \Ints^{N-1}} g(y+e_i) - g(y) +\\
 &\mu_i \sum_{y \in \Ints^{N-1}} (g(y-e_i) - g(y))
  \phi_i(Z_1(t),y)\Biggr) \vartheta(t,y) =0.
\end{align*}
We deduce immediately that
$$
\int_{\bS}
\Omega^{Z_1(t)}(g)(y)\vartheta(t,\diff y)=0
$$
where
$\Omega^{Z_1(t)}$ is the infinitesimal generator of
$(U^{Z_1(t)}(s))$.
This proves exactly that $\vartheta(t,.)$ is invariant for
$U^{Z_1(t)}$. By uniqueness of the invariant distribution of $(U^{Z_1(t)}(s))$,
this implies that, given $Z_1$, $\vartheta(t, \cdot)$ is a deterministic measure for all $t$.
We can deduce that, if $(Y^{K_l})$ is a converging subsequence, then
$(v^{K_l})$ is also converging and its limit is a random measure
in $\mathcal{L}_0(\Ints^{N-1})$. This implies in particular that $(v^{K_l})$
is tight in $\mathcal{L}_0(\Ints^{N-1})$. We can now proceed of the last part
of this step.

We consider $\varepsilon>0$, $\eta>0$ and $t\geq0$.
Because the sequence $(\nu^{K_l})$ is tight in $\mathcal{L}_0(\Ints^{N-1})$, there exists $\kappa>0$ and a
compact $\Gamma \subset \Ints^{N-1}$ such that:
$$
\Pr\left( \sup_{l \geq \kappa} \nu^{K_l}([0,t)\times \Gamma^c) \geq \varepsilon
\right) \leq \eta/2.
$$

Because $Z_1$ is almost surely continuous and $f$ is Lipschitz-continuous,
we have
$$
\Pr\left(\sup_{l\geq \kappa,y \in \Gamma, s\geq t}
  \Bigl|f(Y^{K_l}_1(t),y) -
  f(Z_1(t),y)\Bigr| \geq \varepsilon \right) \leq
\eta/2.
$$
Since $f$ is bounded, we can deduce:
\begin{align*}
\Pr \Biggl( \sup_{k\geq \kappa} \biggl | &\int_{[0,t]\times\bS}
    f(Y_1^{K_l}(s),y) \nu^{K_l}(\diff s\times \diff y)\\
    &-
    \int_{[0,t]\times\bS} f(Z_1(s),y) \nu(\diff s\times
    \diff y) \biggr | \geq 2\varepsilon \|f\| \Biggr) \leq \eta.
\end{align*}

According to \eqref{eq:vartheta}, there exists a family $(\vartheta(t,.))$
of random measures on $\bS$ such that
$$
\sup_{0\leq s \leq t} \left| \int_0^s f(Y_1^{K_l}(u),X_i^{K_l}(K_l u))
-\sum_{y \in\bS} f\left(Z_1(u),y\right)\vartheta(u,y)\diff u \right|
$$
converges in probability to 0 when $K_l$ tends to infinity.

Since $f$ is bounded, we can apply the dominated convergence theorem and we have that
\begin{align*}
\lim_{K_l \to \infty}\mean \Biggl( \sup_{0\leq s \leq t} \Biggl| &\int_0^s f\left(Y_1^{K_l}(u),X_i^{K_l}(K_l u)\right)\\
&-\sum_{y \in\bS} f\left(Z_1(u),y\right)\vartheta(u,y)\diff u \Biggr| \Biggr) = 0.
\end{align*}
We further have that
\begin{align*}
\lim_{K \to \infty}\mean \Biggl( \sup_{0\leq s \leq t} \Biggl| &\int_0^s f\left(Y_1^{K}(u),X_i^{K}(K u)\right)\\
&- \mean\left( f\left(Z_1(u),U_i^{Z_1(u)}(u)\right) \;\middle|\; Z_1(u) \right) \diff u \Biggr| \Biggr) = 0.
\end{align*}
Step 2 is complete.

\subsubsection*{Step 3:}

Using the martingale decomposition of $X_1^K$,
\begin{align*}
{X_1^K( K t) \over K} = &x_1+M_K(t)+{ 1 \over K}\int_{0}^{Kt} \lambda^K_1(s)\diff s\\
&-  {1 \over K}\int_{0}^{ Kt}  \phi_1\left({X_1^K(s)\over K},\ldots, X^N_i(s)\right) \diff s .
\end{align*}
As already remarked in step 1, since $\phi$ is bounded
it follows that $$\mean \left(  {M_K(t)}^2  \right) \leq \frac{A t}{K}$$
which implies using Doob's inequality that there exists a constant $A'$ such that for $K$ big enough:
$$\mean \left( \left| \sup_{0\le s \le t} M_K(s) \right|  \right)  \le  A' \sqrt{t \over K} \le \epsilon.$$

Using the convergence of the arrival process together with the convergence of the martingale $M_K$, we obtain the uniform integrability of $Y^K_1$. (The tightness of $Y_1^K$ has already been obtained in step 1). Now consider a converging subsequence $Y_1^{K_l}$ towards $Z_1$.
Using the results of step 2,  the convergence of the arrival process together and the convergence of the martingale $M_K$, we obtain that $Z_1$ must satisfy:
$$
Z_1(t) = x_1+0 +a_1(t)-  \int_{0}^{ t}  \bar \phi_1(Z_1(s)) \diff s .
$$
Hence the limit is unique and deterministic.

This in turn shows the convergence of $Y_1^K$ in distribution and completely characterize the measure $\vartheta$ introduced in Step 2 as a deterministic measure.
We can now prove the convergence in $L^1$.
Let $\epsilon$ be given.
Define the error estimate:
$$n_K(t)=\sup_{0 \le s \le t}|Y_1^K(s) - u_1(s)|.$$
Define now the noise amplitude as:
$$\bar M_K(t)=\sup_{0 \le s \le t}\left|M_K(s)\right|.$$
Using the convergence of the intensity of the arrival process,
\begin{align*}
n_K(t) \le &\bar M_K(t)+ \epsilon\\
&+\sup_{s \le t} \left\lvert{1 \over K}
\int_{0}^{Ks}\phi_1\left({X_1^K(z) \over K}, X^K_i(z)\right) \diff z -\int_{0}^{s}
\bar \phi_1(u(z))\diff z \right\rvert.
\end{align*}
Using step 2, $\phi_1$ being Lipschitz and bounded, for $K$ large enough:
$$
\mean \left( \sup_{s \le t} \left\lvert{1 \over K} \int_{0}^{Ks}\phi_1\left({X_1^K(z) \over K}, X^K_i(z)\right)\diff z  -\int_{0}^{s} \bar \phi_1(u(z))\diff z \right\rvert \right) \le \epsilon,
$$
which concludes the proof for the $L^1$ convergence of $Y^K_1$.

\section{Qualitative behaviors of the limiting processes}\label{sec:qual}

We describe thereafter the different qualitative behaviors that may occur
depending on the traffic conditions.
Assume that class-$i$, $i=1\ldots,c$ have entered a traffic surge .
Under the scaling considered in Theorem \ref{theo:thetheo}, we observe three qualitative types
of behaviors for the network responses, which are completely characterized
using the stationary distributions of the family of processes
$U_i^x, i=2,\ldots, N$. Defining
\begin{equation}
\forall x\in\R_+^K,\ \forall i \in\{1,\dots,K\},\ \bar \delta_i(x)= \lambda_i-\mu_i\bar\phi_i(x),
\label{eq:delta}
\end{equation}
let $\A$ the set of positive solutions of the equation
$$\bar \delta(x)=0.$$
Given the classical results on asymptotic stability of non-linear autonomous systems,
we can partially classify the possible situations using in particular  the Hartman-Grobman theorem (see for instance \citep{hartman}).
For a $C^1$ flow $\delta$, we write $D\delta(x)<0$ if the linearization of $\delta$
has only eigenvalues with strictly negative real parts and no eigenvalue on the unit complex circle. Assume in the following $\bar \delta$ is $C^1$. We have the possible behaviors:

\begin{enumerate}

\item
The network continues to see class-$i$, $i>c$ saturated (at a macroscopic time and space scales), even after any large (macroscopic) amount of time. A sufficient condition is that there exists $x \in \A$ such that $D\bar \delta(x) <0$ and $x>0$, with initial conditions sufficiently close to $x$. Then the differential equation is asymptotically stable with stable point $x>0$. In this case, limits in time and $K$ cannot commute since there is always a part of the bandwidth of the network used for surging classes, while taking first the limit in time and then the limit in $K$ always converge to the system with allocation $\phi_i(0,\dots,0,x_{c+1},\ldots,x_N)$ for stable classes.

\item The differential equation (\ref{eq:diff}) governing the dynamic of $u$ is unstable, which means that the traffic surge cannot be absorbed and keeps building up. It might lead to the instability of
 other classes in the network.

\item
The traffic surge will be absorbed at macroscopic time, i.e., the differential equation is asymptotically stable with stable point $0$.
Necessary conditions for this situation are that:
\begin{enumerate}
\item $0 \in \A$
\item $D \bar \delta(0)<0$.
\item the initial condition is close enough to $0$, or $\A=\{0\}$.
%\item for all $x \neq 0, ~ x \in \A$, $\delta'(x)>0$.
\end{enumerate}
In this case, note that the stationary measure of $(Y_i^K(t))_{i>c}$ converges when $K \to \infty$ to the  stationary measure of the original system with allocation $\phi_i(0,\dots,0,x_{c+1},\ldots,x_N)$), which boils down to the fact that the limit in time and in $K$ commute for classes $c+1$ to $N$.

\item The traffic surge will be absorbed at macroscopic time, i.e., there exists
a $T>0$, such that for all $t\geq T$, the solution is of the differential equation
is null.

\end{enumerate}

\subsection{Robust bandwidth sharing networks}

Bandwidth sharing networks constitute a natural extension of a
multi-class processor sharing queue, and have become a standard
stochastic model for the flow level dynamics of Internet congestion
control (they were introduced by \citep{massrob}).

\begin{figure}[ht]
\caption{Tree network and linear network}
\begin{minipage}{0.45\linewidth}
\centering\includegraphics[width=0.7\linewidth]{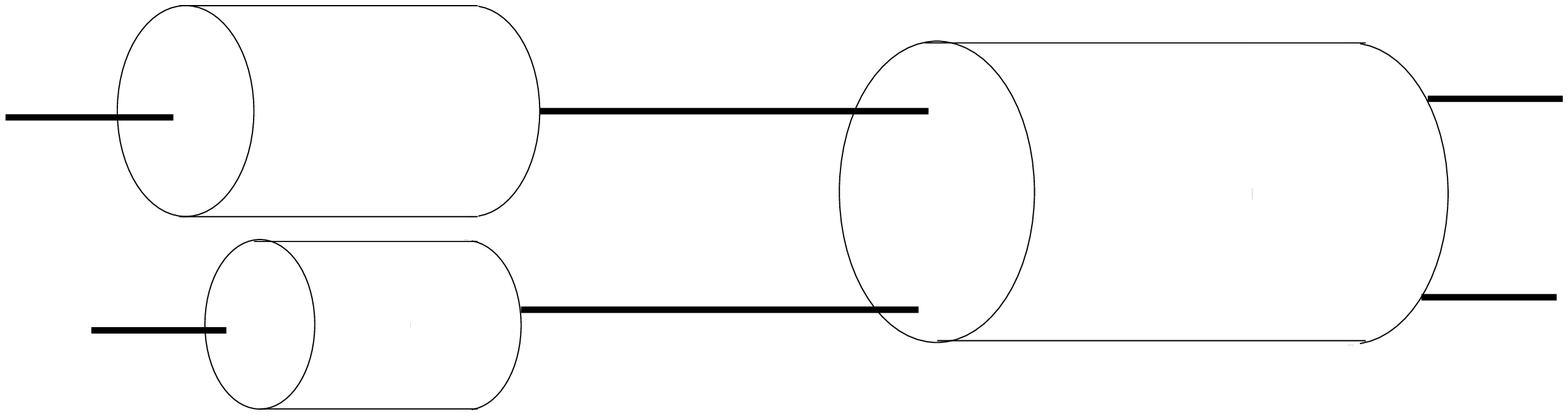}
\end{minipage}
\begin{minipage}{0.45\linewidth}
\centering\includegraphics[width=0.9\linewidth]{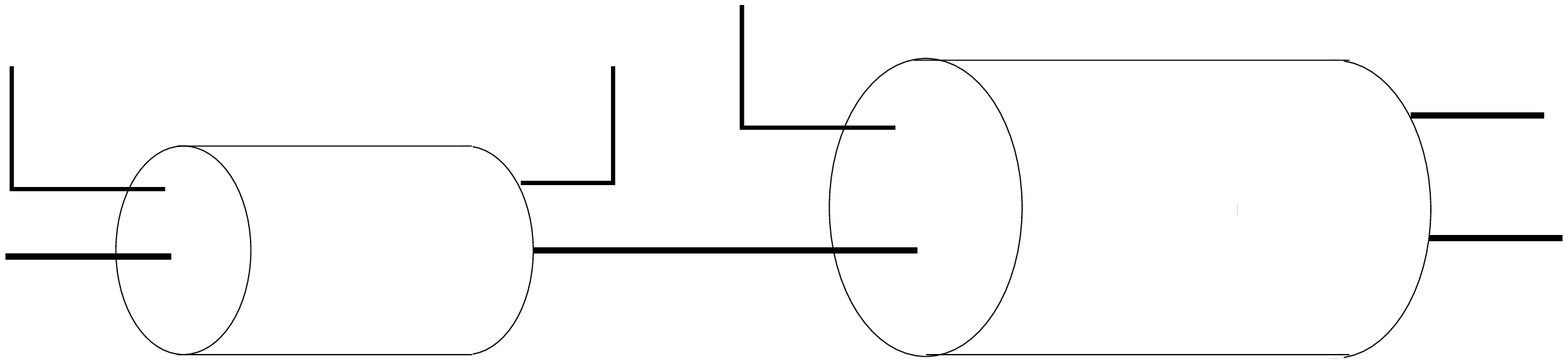}
\end{minipage}
\label{fig:linearnetwork}
\end{figure}

Consider for example the tree network represented on the left of Figure
\ref{fig:linearnetwork}, with two traffic routes, each passing through
a dedicated link, followed by a common link. If each dedicated link
has a capacity $c_i\leq1$, $i=1,\,2$, and the common link has
capacity $1$, the flow on each route gets a capacity $\phi_i(x)$
that lies in the polyhedron $\C$:
\begin{align}
\sum_{i=1}^2\phi_i(x) &\le 1,\\
\phi_i(x) &\le c_i,\quad i=1,2.
\end{align}

%\begin{figure}[ht]
%\caption{Tree network}
%\centering\includegraphics[width=0.4\linewidth]{figs/tree2classes.eps}
%\label{fig:treenetwork}
%\end{figure}

Another example of interest is the linear network represented on the right of
Figure \ref{fig:linearnetwork} with $3$ routes sharing two links.
While the first route passes through both links, routes $2$ and $3$
only use one of the links (one each). This gives the following
capacity constraints:
\begin{align}
\phi_1(x) + \phi_2(x) &\le c_1,\\
\phi_1(x) + \phi_3(x) &\le c_2.
\end{align}

In general, like for the specific foregoing examples, the capacity
constraints determine the space over which a network controller can
choose a desired allocation function. It has been argued by
\citep{kelly1998} that a good approximation of current congestion
control algorithms such as TCP (the Internet's predominant protocol for
controlling congestion) can be obtained by using the weighted proportional
fair allocation, which solves an optimization problem for each
vector $x$ of instantaneous numbers of flows. Specifically, the
weighted proportional fair allocation $\eta(x)$ for state vector $x$
maximizes
$$\sum_{i=1}^N  w_i x_i \log(\eta_i), \eta \in \C, $$
where the weights $w_i$ are class-dependent control parameters.

\begin{remark}
By definition of this optimization program, if
$\phi(\cdot)=\eta(\cdot)$ is the standard (unweighed) proportional
fair allocation with $w_i\equiv1$, then the allocation
$\phi^r(x)=\phi(r.x)$ corresponds to the weighted proportional fair
allocation with weights $w_i\equiv r_i$.
\end{remark}

This framework has been generalized to so-called weighted
$\alpha$-fair allocations, which provide flexibility to model
different levels of fairness in the network. Another important
alternative is the balanced fair allocation \citep{BFreview}, which
allows a closed form expression for the stationary distribution of
the numbers of flows in progress. In addition, the balanced fair
allocation gives a good approximation of the proportional fair
allocation while being easily evaluated, which is attractive for
performance evaluation.

Remind that all $\alpha$-fair bandwidth sharing are stable for $\alpha >0$ (in the sense that the process $X$ is positive recurrent)
\citep{bonald-massoulie-01,de-veciana-01} if
$$\rho \in \S=\{ \eta, A \eta \le C\}.$$

We now refine the concept of stability by saying that the network is stable if classes undergoing a surge eventually drain while the other classes stay stochastically stable.
 More formally, we say that
\begin{definition}
 The network is robust stable if:
$$
 \limsup_{t \to \infty}\limsup_{|x| \to \infty} {E^{x}[X^{r}(|x|t)_k] \over |x|}=0.
$$
\end{definition}

Define further the robust stability region as the set of parameters such that the network is robust stable, i.e.:
\begin{equation}
\S^r =\left\{ \rho \in \mathbb R^N_+ :  ~ \limsup_{t \to \infty}\sup_{|x| \to \infty}  {E^{x}[X^r_k(|x|t)] \over |x|} =0 \right\}.
\end{equation}
% Remark, that as previously underlined,  when all classes are undergoing a surge of traffic, we retrieve the usual notion of fluid limit, for which the convergence to $0$ implies the network (usual notion of) stability. Hence it holds in general that:
% $$\S^r \subset \S.$$

In the sequel, we first show that for work-conserving allocations, the robust stability set coincides with the usual stability set (except possibly on a negligible set of parameters).
\newline On the other hand, the situation is much more complex for non-work-conserving allocations where even with appropriate weighted allocations, surges might not be asymptotically transparent to the other classes while classes undergoing surges might get asymptotically strictly more bandwidth than in the case of the allocation giving full priority to stable classes. We give an example of this phenomenon later on. %(see for instance Figure \ref{fig:dich_tree3} in the examples below).

For monotonic networks, we however prove that the robust stability coincide with
the set of parameters under which a ``priority allocation'' is stable.

%If the allocation giving full priority
%to stable classes i.e., the allocation defined by
%\begin{align*}
%\phi_1^{PR}(x)&=\phi_1(x_1,\ldots x_c,0,\ldots,0)\ind_{\{x_{c+1}=0,\ldots,x_N=0\}},\\
%\phi_i^{PR}(x)&=\phi_i(0,x_{c+1},\ldots,x_N), ~ \forall i\ge c+1
%\end{align*}
%is stable, then the differential equation (\ref{eq:diff}) has the stable point $0$.
%This is consistent with the fact that the priority allocation constitutes a worst case scenario for class-$1$.
%Necessary and sufficient conditions of stability stay a challenging direction future research.
%We now give more details of these findings. For the ease of exposition,  in this section, we suppose
%that the arrival rate of class-$1$ is fixed.

\subsection{Existence of the fluid limit}

\begin{lemma}
If $\rho \in \S$, then the processes $U^{z}$ are positive recurrent for any $z$,
and Theorem \ref{theo:thetheo} applies.
\end{lemma}

\bp
Asymptotically, when $x_i \to \infty, i >c$ the allocation allocated to classes $i>c$
coincides with the allocation $\phi(0,\ldots,0, x_{c+1}, \ldots, x_N)$ which is stable
under the usual conditions of traffic.
\ep

\subsection{Work-conserving allocations}\label{subsubsec:WC}

Consider a work conserving allocation such that
$$\forall x \neq 0, ~ \sum_{i=1}^N\phi_i(x)=1.$$
Every work-conserving allocation has the same stability region, namely $$\sum_{i=1}^N \rho_i < 1.$$ If the (usual) stability condition is satisfied, then the priority mechanism considered
is asymptotically  equivalent to giving full priority to class-$i, ~i\ge c+1$.
In other words, the fluid limit obtained for class-$1$ is in that case the same as the fluid limit of an allocation that gives a full priority to class-$i,~i \ge c+1$, which we prove in the following Proposition.
\begin{proposition}\label{prop:WC}
For a work conserving network, $\S^r=\S$ (except possibly on the frontier of the stability sets) and:
$$Y^K_i(t) \stackrel{L^1}{\to}u_i(t)=\left(u_i(0)+ \lambda_i-\mu_i\left(1-\sum_{i=c+1}^N\rho_i\right)t\right)^+.$$
\end{proposition}
\proof
Assume $\sum_{i=1}^N \rho_i < 1$ in which case the network is stable.
Fix $z_1 \in \mathbb R$.
Using the conservation of the rates at equilibrium for the process $U^{z_1}$ ( which boils down in the Markovian context
to saying that at equilibrium the drift of $y \to y_i$ should be $0$), we can write that:
$$
\sum_{i=c+1}^N\sum_y \phi_i(z_1,y)\pi^{z_1}(y)=\sum_{i=c}^N \rho_i.
$$
We now calculate $\bar \phi_i$ for $z_1>0$:
\begin{align*}
\bar \phi_i(x) &= \sum_y    \phi_i(z_1, y) \pi^{z_1}(y),\\
\bar \phi_i(x) &= \sum_y   (1- \sum_{j\ge c+1}\phi_j(z_1,y)) \pi^{z_1}(y),\\
\bar \phi_i(x)&= 1- \sum_{j\ge c+1}\rho_j.
\end{align*}
Hence, the capacity seen asymptotically by class-$1$ is ($1-\sum_{j\ge c+1}\rho_j$), which concludes the proof.

\paragraph{Example: one link with the DPS allocation}

The simplest instance of a network consists of one link
shared by several competing classes of traffic.
If the initial policy is supposed to be the classical processor sharing policy:
$\phi_i(x)={x_i \over |x|}$
then the prioritized version of the model becomes the so-called discriminatory processor sharing (DPS): $\phi_i(r.x)={r_i x_i \over \sum_j r_j x_j}.$

\begin{figure}
      \caption{DPS with three classes: scaling of \mbox{class-$1$} (left) and of \mbox{class-$2$} (right)}
      \label{fig:dps3c_1}
        \centering\includegraphics[width=0.4\linewidth]{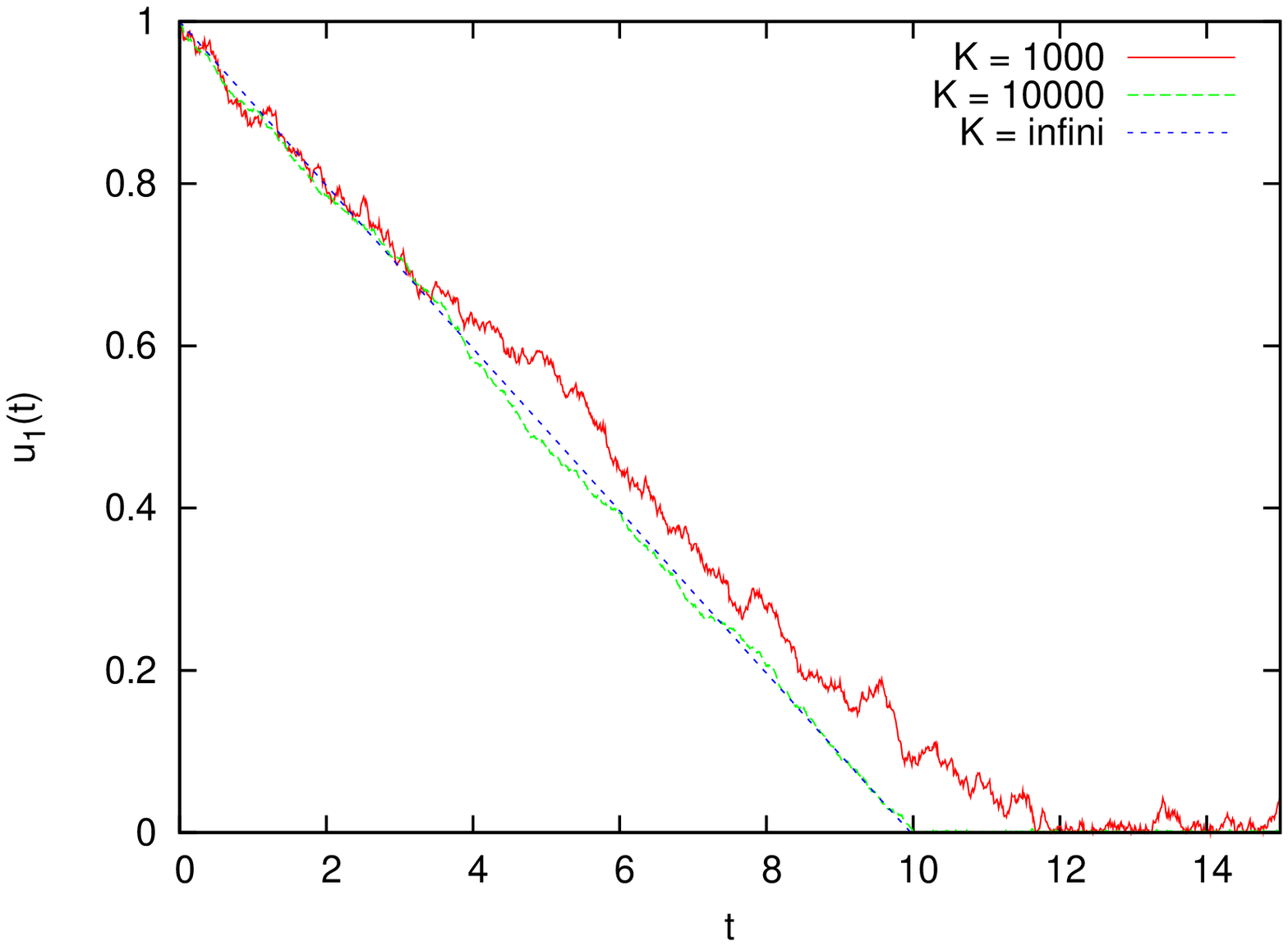}
        \includegraphics[width=0.4\linewidth]{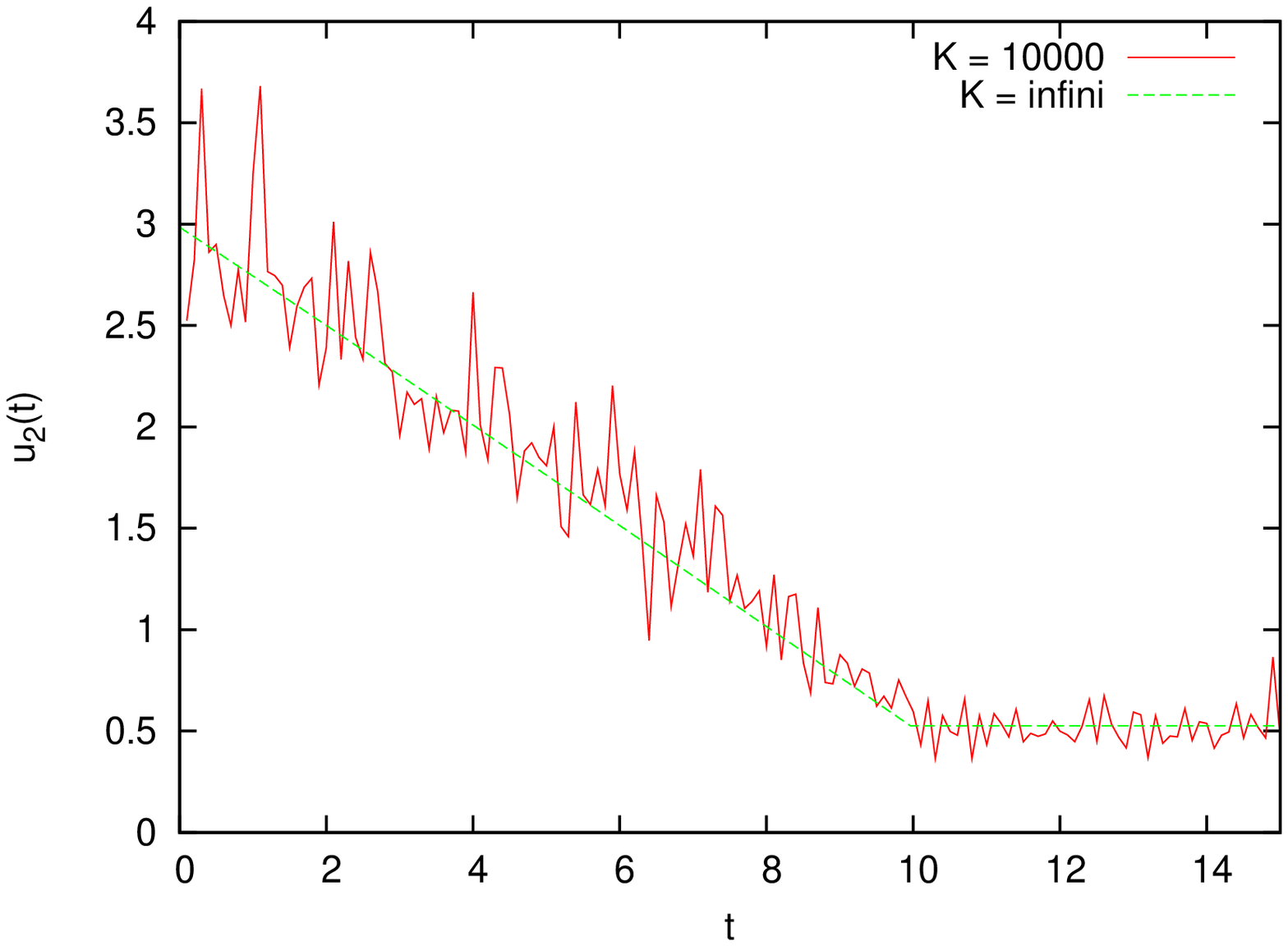}
    \end{figure}

Consider a single link of capacity $1$ shared
by three classes. The bandwidth is allocated according to DPS with weight
$r_i$ for class $i$, $i=1,2,3$. Proposition \ref{prop:WC} says that $u_1(t)$
is a straight line with slope $\lambda_1 -\mu_1(1-(\rho_2 + \rho_3))$.
This behavior is illustrated in Figure \ref{fig:dps3c_1}, for which
$\lambda_1 = 0.5, \mu_1 = 1, \rho_2 = 0.3, \rho_3 = 0.1$. The slope
calculated using the proposition is thus $0.1$, which is verified
in the figure.

In Figure \ref{fig:dps3c_1}, we plot the empirical mean of class-$2$ at a macroscopic scale,
(i.e. ${1 \over s}\int_{t}^{t+s} f(Y^K(h))\diff h$)  for a temporal window of $s=0.1$.
%Note that the temporal window being relatively

\subsection{Monotone allocations}

Define the allocation $\psi$ giving full priority to class $c+1$ to $N$, given by
\begin{eqnarray*}
\psi(x)&=&\phi(0,\ldots,0,x_{c+1},\ldots, x_N), \ \mbox{if  } x_i >0, \text{for some } i>c.\\
\psi(x)&=&\phi(x), \text{ otherwise.}
\end{eqnarray*}
Denote $\S(\psi)$ the stability region of the network with allocation $\psi$.

\begin{proposition}\label{prop:NWC}
Consider a monotonic allocation (i.e. such that $\phi_i$ is decreasing in $x_j$, $j \neq i$).

If $\bar\delta(0)<0$ where $\bar\delta(x)$ is defined by \eqref{eq:delta}, then the network is robust stable and surging classes do not influence asymptotically stable classes.
 Conversely if $\bar\delta(0)>0$, the network is not robust stable.

Moreover $\bar \delta(0) <0$ if the network with allocation $\psi$ is stable i.e.:
$$\S^r(\phi)=\S(\psi).$$
\end{proposition}

\proof
Using stochastic comparisons (see \citep{stabST} for more
details on stochastic comparisons of multidimensional birth-and-death processes with
monotonic allocations), we obtain that
$$ U^{0}_i \le_{\mathrm{st}} U^{z} , \forall z ,$$
which implies that $\forall z, ~ \bar \phi_i(z) \ge \bar \phi_i(0),$
This in turn implies that
$$
{d \over dt} u_i(t)  < \bar{\delta}(0) <0,\quad \forall t \geq0,\ \text{such that }u_i(t)>0.
$$
This implies that $u_i$ will reach 0 in finite time.

The reverse statement follows along the same lines.
\ep

\paragraph{Example: a tree network}\label{sec:numtree}

    Let us consider the tree network shown in Figure \ref{fig:linearnetwork} with $c_1 = 0.4$ and
        $c_2 = 0.8$. We shall assume the following bandwidth allocation:
        Define $\mathcal{S}_1 = \{(x_1,x_2): (r_1x_1 + r_2x_2)c_1 < r_1x_1\}$.
        For $x_1 > 0$ and $x_2 > 0$,
        \begin{equation}
        \phi_1(x_1,x_2) = \begin{cases}
            c_1, & \text{if }(x_1,x_2) \in \S_1, \\
            \max\left(\frac{r_1x_1}{r_1x_1+r_2x_2}, 1 - c_2\right), &\text{if } (x_1,x_2) \in \S_1^c,
        \end{cases}
        \end{equation}
        and $\phi_2 = 1 - \phi_1$.

\begin{figure}[h]
                \caption{Partitioning of the stability region for the tree network}
                \label{fig:stabreg_tree}
                \centering\includegraphics[width=6cm]{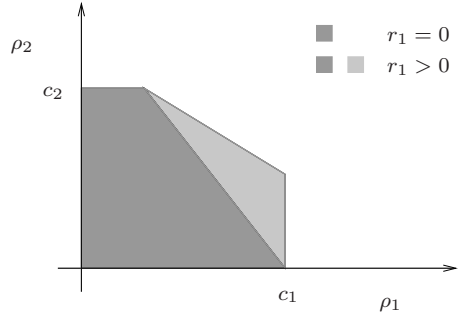}
        \end{figure}

        For this network, the allocation becomes a strict priority allocation for
        class-$2$ when $r_1 = 0$, in which case class-$1$ gets capacity $c_1$ if there are no class-$2$ flows, and $1-c_2$
        otherwise. Thus, for a fixed value of $\rho_2$, class-$1$ is stable if
        $\rho_1 < \left(1-\frac{\rho_2}{c_2}\right)c_1 + \frac{\rho_2}{c_2}(1-c_2)$.
        The stability regions for $r_1 = 0$  and $r_1 > 0$ are shown in Figure \ref{fig:stabreg_tree}.

\begin{figure}[ht]
        \caption{Tree network: scaling of class-$1$ (left) and of class-$2$ (right)}
        \label{fig:dich_tree1}
        \includegraphics[width=6cm]{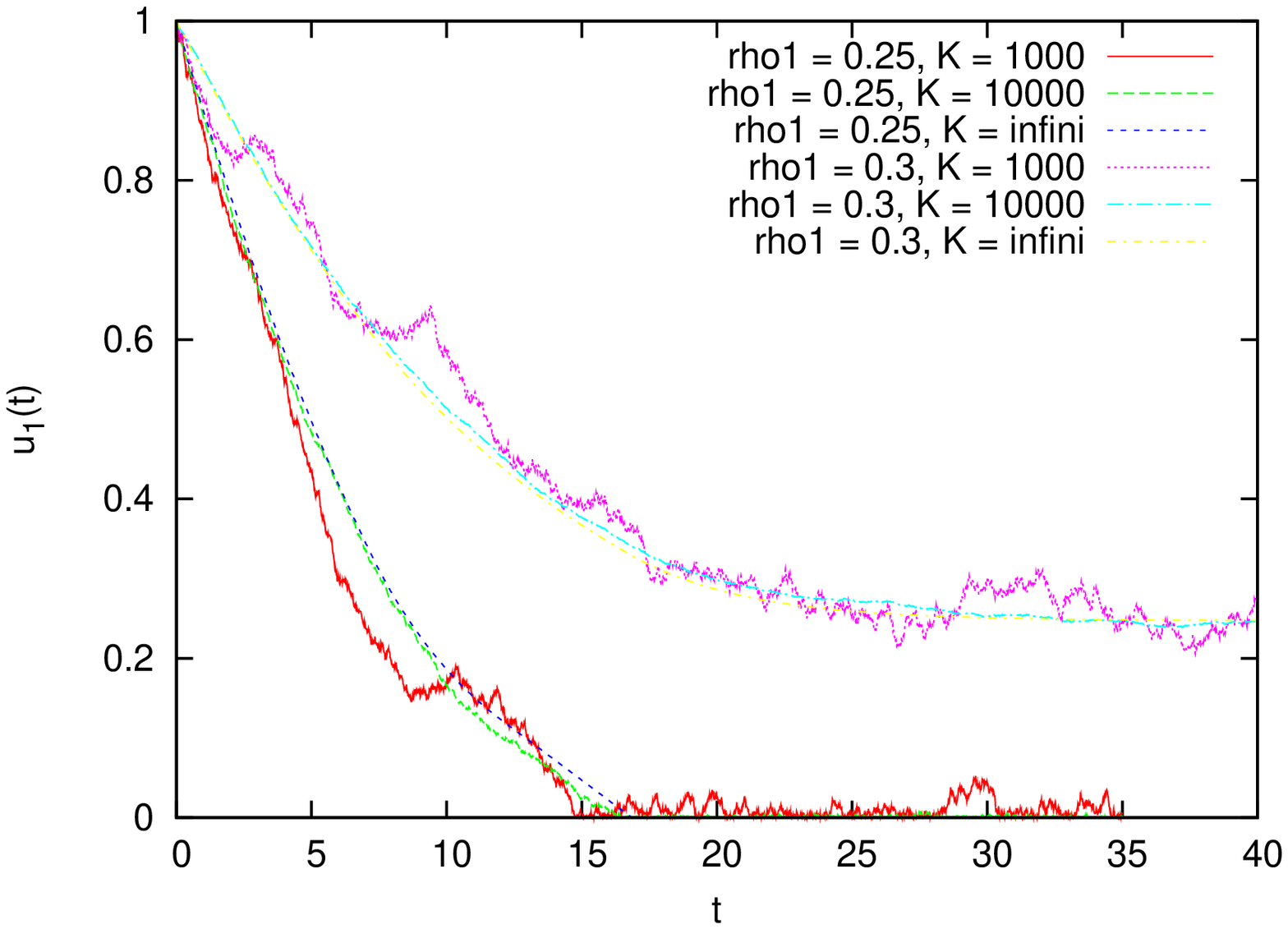}
	\psfragscanon
        \centering\includegraphics[width=6cm]{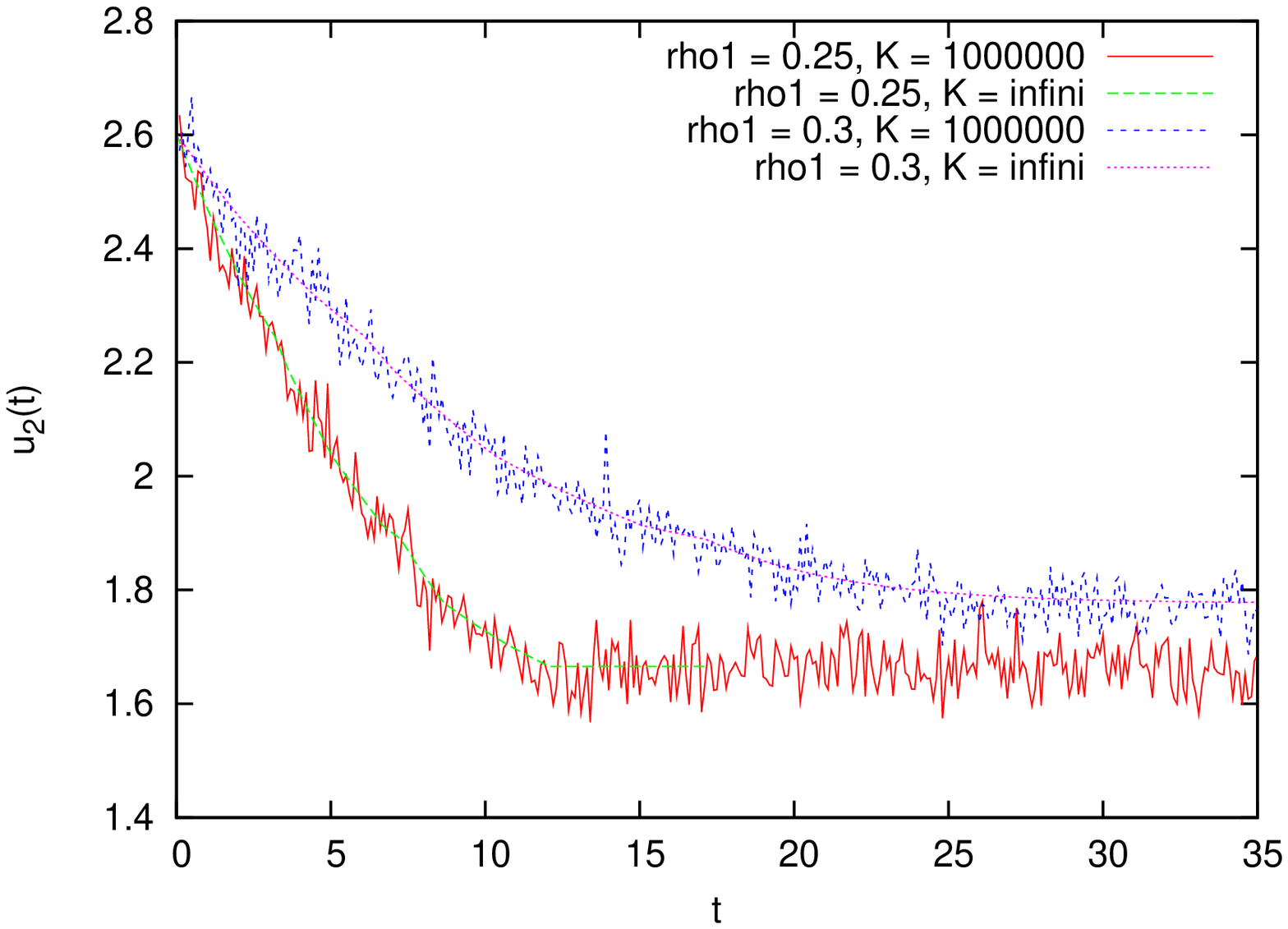}
        \end{figure}

        The dynamics of $u_1(t)$ for two different values of $\rho_1$ -- one in each region -- is plotted in
        Figure \ref{fig:dich_tree1}, for which $\rho_2 = 0.5$.

        For class $2$, when the priority allocation is stable the dynamics of the average number of customers
    converges to the one of the priority allocation, that is $\rho_2/(c_2-\rho_2)$, as is
    illustrated in Figure \ref{fig:dich_tree1}.

\begin{figure}[ht]
   \caption{Tree network: comparisons of trajectories of class-$1$ for a proportional fair allocation and a priority (to class-$2$) allocation}
   \label{fig:dich_tree3}
   \centering\includegraphics[width=7cm]{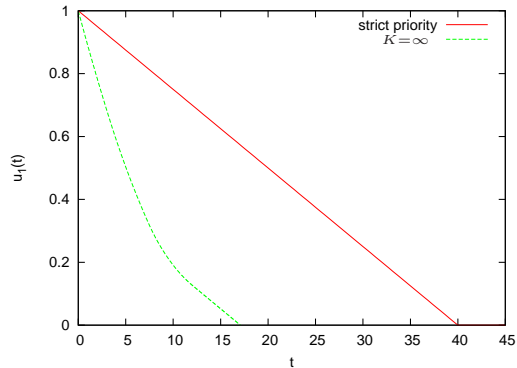}
 \end{figure}

In Figure \ref{fig:dich_tree3}, we show how class-$1$ is actually favored by asymptotically using the bandwidth of class-$2$, compared to the case
where class-$2$ is given a strict priority.

\subsection{Non-monotone networks}

For non-monotonic networks, some unusual behaviors can be observed; for instance
the usual fluid limit may exhibit the following behavior: one class of traffic can reach $0$
at the fluid scale and stay at 0 for a finite time before increasing again. We here
show on an example how the priority scaling avoids this kind of behavior.

Here, we consider a linear network with two links and
three classes of flows as shown in Figure \ref{fig:linearnetwork}.

    Let $\alpha_i$ be the capacity allocated to a flow of class-$i$.
    The capacity allocated to class-$i$, $\phi_i$, is then $x_i\alpha_i$. The bandwidth is allocated according to the
    weighted proportional fair allocation, that is, $(\alpha_1, \alpha_2, \alpha_3)$ is the solution of the
    following maximization problem:
    \begin{equation*}
    \begin{aligned}
      &\text{maximize} && r_1x_1\log(\alpha_1) + r_2x_2\log(\alpha_2) + r_3x_3\log(\alpha_3) \\
      &\text{subject to} && x_1\alpha_1 + x_2\alpha_2 \leq c_1, \\
            &&& x_1\alpha_1 + x_3\alpha_3 \leq c_2,\\
    \end{aligned}
    \end{equation*}
    where $r_i$ is the weight of class $i$.

\begin{figure}[h]
\caption{Usual fluid (left) and priority scaling (right) for the linear network.}
\vspace{5mm}
\centering\includegraphics[width=0.4\linewidth]{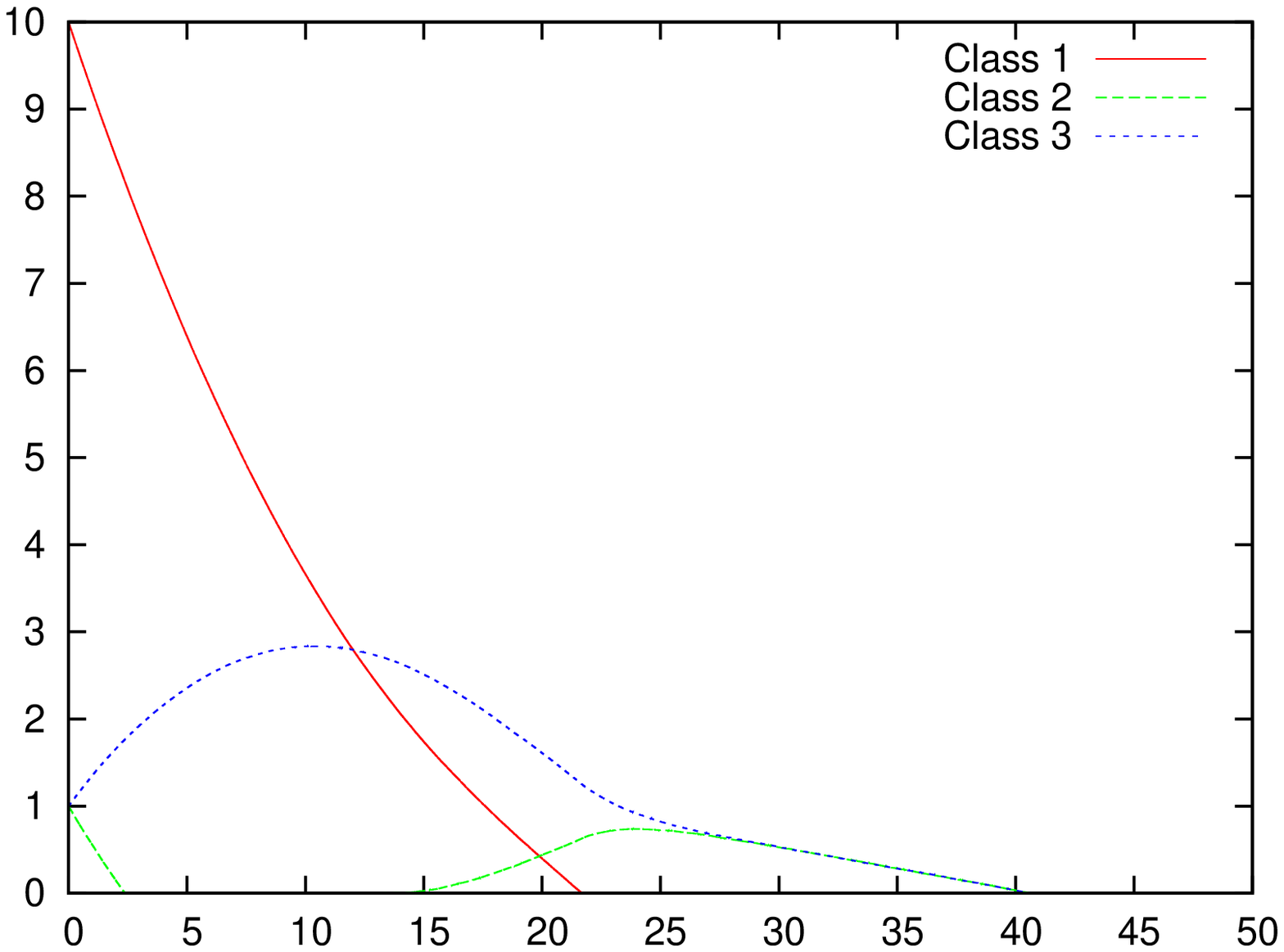}
\includegraphics[width=0.4\linewidth]{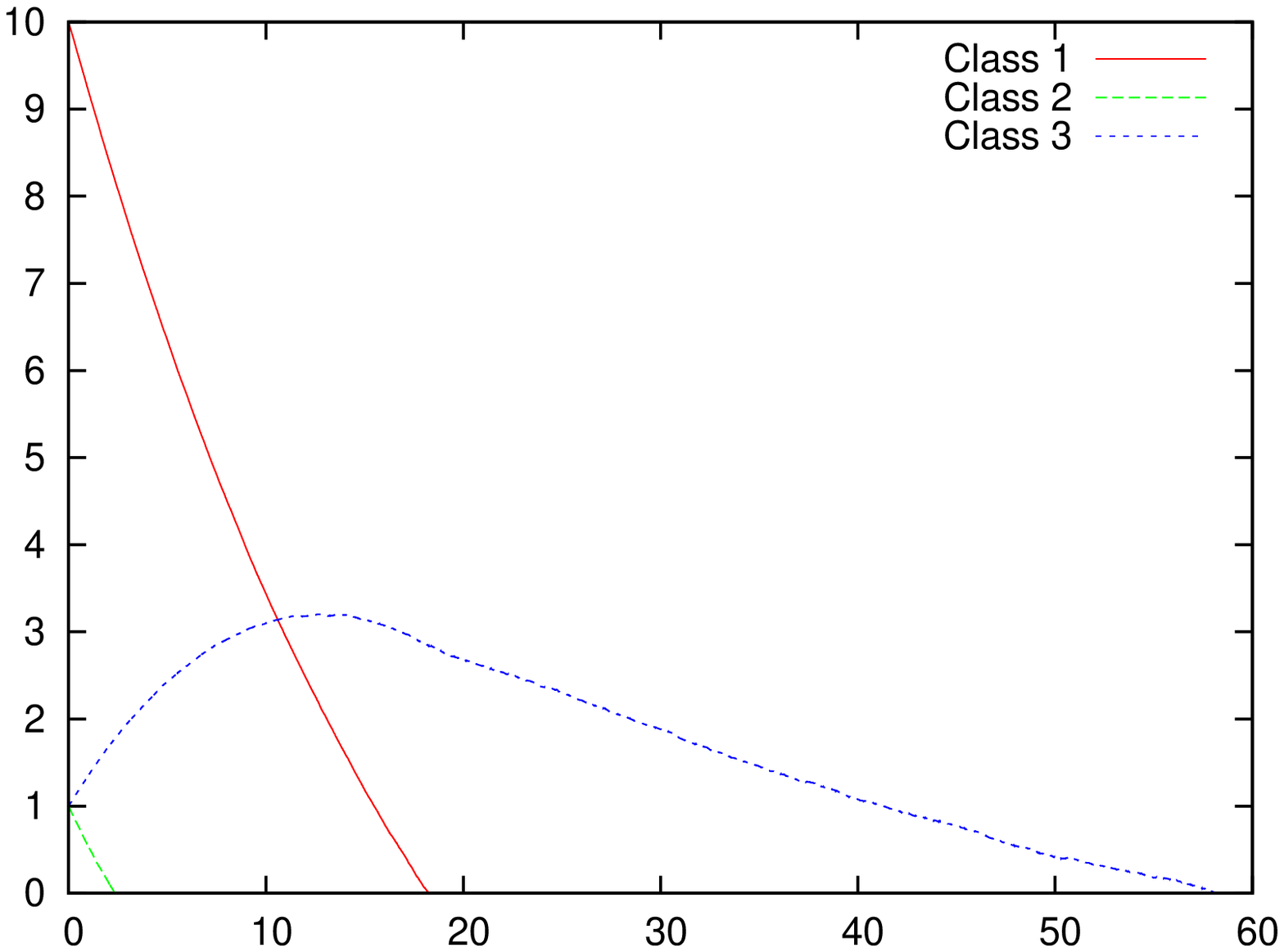}
\label{fig:lin_net_sim}
\end{figure}

Consider the following network parameters with
arrival rates and service rates:
\begin{align*}
    c_1 = c_2 = 1, \; r_1 = r_2 = r_3 = 1, \\
    X_1(0) = 10\cdot K, X_2(0) = K, X_3(0) = K,
\end{align*}
that is, the three classes are unstable at the beginning. The trajectories of the number of flows as
a function of the scaled time for $K=10000$ is shown in Figure \ref{fig:lin_net_sim}.

Concerning the usual fluid limit, we observe that Class $2$ becomes stable around
the $2$ time unit mark. It then becomes unstable
around the $15$ time unit mark, and becomes stable again around the $40$ time unit mark.
This behavior can be explained as follows.

In the first part of the trajectory, class $2$ gets sufficient capacity to drain out
while the number of class $3$ flows grows.
In the second part, since Link $2$ is not a bottleneck for Class $3$ flows and Class $1$ gets smaller,
Class $3$ gets a larger share on Link $2$. Meanwhile, the arrival and service
rates of Class $2$ and Class $3$ being the same, the imbalance in the rate allocation
means that Class $2$ is now unstable and its number of flows starts to grow until it reaches the same number
as that of Class $3$ flows at which time they both share the link capacity equally. Since the network is
stable, all the three classes drain out eventually.

Due to the priority mechanism employed to penalize unstable classes, such a phenomenon
does not happen in the case of the priority scaling as illustrated in the right-hand
Figure.

\section{Integration of streaming and elastic traffic}\label{sec:integration}

Consider now a system where two intrinsically different types of
traffic -- ``streaming'' and ``elastic'' traffic -- coexist and
share a given link. Such models have been considered by
\citep{sindo,decoigne,bonaldsig2004}. It is natural to equip
streaming traffic with a fixed required rate, say, $c$ per flow.
Giving priority to streaming traffic (class 2)
the allocation of service may be chosen as:
\begin{align*}
\phi_1(x)&=\max\left({r_1 x_1 \over r_1 x_1 + c x_2},1-cx_2\right),\\
\phi_2(x)&=  c x_2,
\end{align*}
where the parameter $r_1$ quantifies the level of priority. The
allocated capacity cannot exceed the total capacity. If the latter
is normalized to 1, the state space must be restricted to states
$x_2$ such that
$$\phi_1(x)+\phi_2(x)\le 1.$$
Then, if the number of current streaming flows $x_2$ is such that
$\phi_1(x+e_2)+\phi_2(x+e_2)> 1,$ arriving streaming flows must be
blocked from the network.

In this example, if accepted in the network, the
 capacity allocation of class-$2$ flows is $c x_2$ independently of the number of flows of class-$1$,
    for all values of $r_1 > 0$, whereas the capacity allocation of class-$1$ flows depends on the
    number of class-$2$ flows such that $\phi_1(x_1, x_2) = r_1x_1/(r_1x_1 + cx_2)$.
    However, class-$2$ flows are admitted only if there is sufficient capacity, that is, if
    ${u_1(t) \over u_1(t) + c(x_2+1)} + c (x_2+1) \le 1$.

    Denote $$\S_{z_1}=\left\{ x_2: {z_1 \over z_1 + c x_2} + c x_2 \le 1\right\},$$
    the state space of class-$2$ conditioned on $u_1(t) = z_1$. Define $\rho_2= {\lambda_2 \over \mu_2 c}$. The process $U^{z_1}_2$ is
    birth-death process with
    birth rate $\lambda_2$ and death rate $\mu_2 c x_2$, and whose stationary distribution is given by
        \[
                \pi_2(x_2) = \frac{1}{\sum_{j\in\S_{z_1}}\rho_2^j/j!}\frac{\rho_2^{x_2}}{x_2!}.
        \]
        For the priority allocation, class-$1$ is stable if and only if $\rho_1 < \pi_2(0)$. Thus,
    if $\rho_1 < \pi_2(0)$, then the limit point of $u_1(t)$ is $0$, and if
    $\pi_2(0) < \rho_1 < 1$, then the limit point is positive.

Performing the scaling previously defined, remark that the
 state space depends for a fixed macroscopic state $z_1$ on both $z_1$ and $c$.
We can apply Theorem \ref{theo:thetheo} with $\bar \phi_1$ being defined by:
$$\bar \phi_1(z)=\sum_{x_2 \in \S_{z_1}}{z_1 \over z_1 + c x_2} {\rho^{x_2} \over x_2!} C(z_1),$$
where $C(z_1)=(\sum_{x_2 \in \S_{z_1}} {\rho^{x_2} \over x_2!})^{-1}$.
In the case that $c$ is very small ($c << 1$), we might consider as a reasonable approximation
a Poisson distribution for class-$2$, whatever the state of class-$1$.
In that case, $\bar \phi_1$ takes a slightly simpler form. After simple calculations:
$$\bar \phi_1(c z_1)=H(z_1)= {z_1\int_{0}^{\rho_2} u^{z_1-1}\exp(u)\diff u \over \rho_2^{z_1} \exp(\rho_2)}.$$
This allows a recursive evaluation for integer-valued $z_1$. Using simple calculus, for $n \in \mathbb N$:
$$H(n+1)={n+1 \over \rho_2} (1-H(n))$$
We can also evaluate $H$ in terms of special functions:
$$H(n)={n!-n\Gamma(n,-1) \over (-\rho_2)^{n} \exp(\rho_2)},$$
where $\Gamma(n,-1)$ is the incomplete $\Gamma$ function.

\begin{figure}[h]
        \caption{Streaming and elastic traffic: scaling of the elastic traffic}
        \label{fig:dich_stream2}
        \centering\includegraphics[width=7cm]{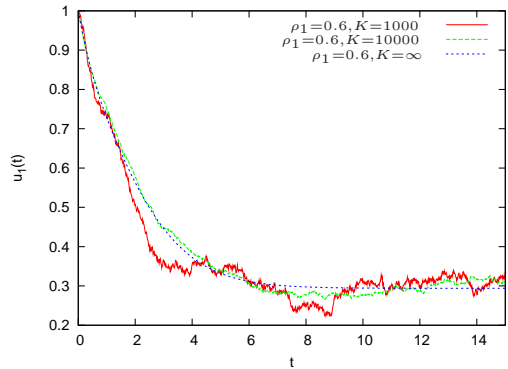}
    \end{figure}

In Figure \ref{fig:dich_stream2}, we plot the $u_1(t)$ for $\rho_1 = 0.6$, $\rho_2 = 0.2$, and $c = 0.01$.

	\subsection{Quality-of-Service guarantee}

	The Quality-of-Service for streaming flows is mainly characterized by the
	probability that an incoming flow does not find sufficient capacity in the
	network for the flow to be accepted. In networks in which streaming and elastic
	traffic do not interact, this probability can be computed using an Erlang
	Fixed-Point approximation \citep{Kelly86}. However, in the context of the present
	example, the interaction of these two types of traffic makes it more difficult
	to apply these fixed-point approximations, mainly due to the fact that the
	state space of the elastic flows is unbounded. However, in the limiting regime
	under consideration, we can come up with a rule-of-thumb that can be used to
	guarantee a blocking probability smaller than a desired value.

	First, we consider a single link whose capacity is shared by the two types of
	flows. Let $p_m$ denote the desired maximal
	blocking probability of class-$2$ flows. We shall set the priority level of
	class-$1$ (by varying $r_1$) such that the probability of blocking of class-$2$
	is always less than $p_m$.

	For $u_1(t) = z_1$, an arrival of class-$2$ is blocked if and only if
	$\frac{z_1}{z_1 + c(x_2 + 1)} + c(x_2 + 1) < 1$, which is equivalent to
	$\frac{1-z_1}{c} - 1 < x_2 \leq \frac{1-z_1}{c}$. The term $\frac{1-z_1}{c}$ is the
	number of circuits of size $c$ available when the total capacity is $1-z_1$. Thus,
	\[
		\tilde{N}_2 = \left\lfloor \frac{1-z_1}{c} \right\rfloor,
	\]
	is the maximum number of simultaneous flows of class-$2$ in the system, and an arrival of class-$2$
	is blocked if and only if the number of flows of class-$2$ is $\tilde{N}_2$.

	Let $g(n)$ denote the blocking probability when the number of circuits is the network is $n$. From the Erlang-B
	formula,
	\[
		g(n) = \frac{\rho_2^{n}/n!}{\sum_{j=0}^{n}\rho_2^j/j!}.
	\]
	The inverse function $g^{-1}(p_m)$ gives the minimum number of circuits required to ensure a blocking
	probability smaller than $p_m$. In order to guarantee a maximal blocking of $p_m$ the number of
	circuits, $\tilde{N}_2$ has to be larger than $g^{-1}(p_m)$ at all instant of time, which leads us to
	the following necessary and sufficient condition for guaranteeing the Quality-of-Service of class-$2$ flows:
	\[
		\bar u_1 := \sup_{0 \leq t < \infty} u_1(t) < 1 - c\lceil g^{-1}(p_m) \rceil.
	\]
	We can ensure the above inequality by scaling the process $u_1(t)$ by a factor
	$\frac{1-c\lceil g^{-1}(p_m)\rceil}{\bar u_1}$. This, in turn, can be achieved by scaling the
	priority level (or, equivalently, $r_1$) by this very same factor. This additional scaling
	results in a larger share of the bandwidth for class-$1$ flows in case $1-c\lceil g^{-1}(p_m)\rceil > \bar u_1$.
	Conversly, if $1-c\lceil g^{-1}(p_m)\rceil < \bar u_1$, the priority level of class-$1$ flows is
	appropriately decreased so that the blocking probability constraint of class-$2$ flows is
	not violated.

	%{\bf Conjecture:}
	Using the monotonicity of $\phi_1$ in its first variable, we get that if
	$\lambda_1 > \bar\phi_1(u_1(0))$, then $u_1(t)$ converges monotonically to its limit point.
	 Hence,
	\[
		\bar u_1 = \begin{cases}
					u_1(0), & \text{if }\lambda_1 < \bar\phi_1(u_1(0));\\
					\bar\phi_1^{-1}(\lambda_1), & \text{otherwise}.
				\end{cases}
	\]
	\begin{remark}
	The blocking probability for a given value of $z_1$ is in fact a conditional blocking
	probability in the sense that it is the fraction of calls dropped when the class-$1$
	flows take away a capacity of $z_1$. The unconditional blocking probability of class-$2$
	flows can be computed by integrating over $z_1$, which is rather conservative. An
	alternative scaling could be constructed such that only the unconditional blocking
	probability satisfies a given constraint.
	\end{remark}

	 \begin{remark}
	In a network of links shared by several classes of streaming flows and one class of
	elastic flow, we could use fixed-point approximations to compute the blocking probability
	for the different classes of streaming flows as a function of $z_1$. Assuming that
	this probability is increasing in $z_1$,  we could then compute the maximum value
	that $z_1$ can attain without the streaming classes violating their individual
	blocking probability.
	\end{remark}

\section{Conclusions}\label{sec:conclusion}
We analyzed the flow-level performance of multi-class
communication networks when one of the classes undergoes
a traffic surge. We showed that, under an appropriate scaling of space and
time, the dynamics of the temporarily unstable
class can be described by a deterministic differential equation in
which the time derivative at a given point
depends on the conditional stationary distribution of the other
classes calculated at that point. For work-conserving
allocations, the differential equation is the same as the one of the
network in which other classes have strict priority
over the temporarily unstable class, that is, the scaled process evolves linearly
 and is either absorbed at zero
or grows indefinitely depending on whether the network is stable or not.

For non-work conserving allocations, the trajectory is much more
complex to describe as it depends on the mean residual bandwidth left over
by the other classes which in turn depends on the current state of the first class.
The limit point of the fluid trajectory can hence be non-zero and finite.
We characterized the robust stability region of monotone allocations.
We illustrated
this behavior through several examples of network topologies and
bandwidth allocations that are commonly used to model communication
networks.

The time-space-transitions scaling that we considered raises several open
questions which would give a better understanding of the network dynamics.
 In particular, finding
necessary and sufficient conditions for the limit point of non
work-conserving allocations to be zero would constitute a very interesting result.
Also, error bounds estimates would be necessary to obtain
a reliable performance evaluation tool. %These bounds cab difficult to obtain since

\subsection*{Acknowledgment}
We would like to thank A. Ferragut and F. Paganini for fruitful discussions on the example 4.5.

\bibliographystyle{imsart-nameyear}
\bibliography{itc2010}

\end{document}